\documentclass[aps,pre,showpacs,floatfix,twocolumn,10pt]{revtex4-1}

\usepackage[T1]{fontenc}

\usepackage{bm}				
\usepackage{amsmath}
\usepackage{amssymb}
\usepackage{latexsym}
\usepackage{amsfonts}
\usepackage{graphicx}
\usepackage{color}
\usepackage{wasysym}

\newcommand{\be}{\begin{equation}}
\newcommand{\ee}{\end{equation}}
\newcommand{\la}{\langle}
\newcommand{\ra}{\rangle}
\newcommand{\ben}{\begin{eqnarray}}
\newcommand{\een}{\end{eqnarray}}

\renewcommand{\vec}[1]{{\bf {#1}}}

\newcommand{\vv}{{\bf{v}}}

\def\(({\left(}
\def\)){\right)}
\def\[[{\left[}
\def\]]{\right]}

\begin{document}

\title{Probing local equilibrium in nonequilibrium fluids}

\author{J.J. del Pozo}
\email{jpozo@onsager.ugr.es}
\author{P.L. Garrido}
\email{garrido@onsager.ugr.es}
\author{P.I. Hurtado}
\email{phurtado@onsager.ugr.es}
\affiliation{Institute Carlos I for Theoretical and Computational Physics, and Departamento de Electromagnetismo y F\'isica de la Materia, Universidad de Granada, 18071 Granada, Spain}

\date{\today}

\pacs{
05.40.-a,		
05.70.Ln,		
74.40.Gh,		
65.20.-w		
}

\begin{abstract}
We use extensive computer simulations to probe local thermodynamic equilibrium (LTE) in a quintessential model fluid, the two-dimensional
hard-disks system. We show that macroscopic LTE is a property much stronger than previously anticipated, even in the presence of important finite size effects,
revealing a remarkable bulk-boundary decoupling phenomenon in fluids out of equilibrium. This allows us to measure the fluid's equation of state in simulations
far from equilibrium, with an excellent accuracy comparable to the best equilibrium simulations. Subtle corrections to LTE are found in the fluctuations of the
total energy which strongly point out to the nonlocality of the nonequilibrium potential governing the fluid's macroscopic behavior out of equilibrium.
\end{abstract}

\maketitle


\section{Introduction}
\label{sec1}

Nonequilibrium phenomena characterize the physics of many natural systems, and their understanding remains as a major challenge of modern theoretical physics.
Out of equilibrium, dynamics and statistics are so intimately knotted that no general bottom-up approach exists yet (similar to equilibrium ensemble theory) 
capable of predicting nonequilibrium macroscopic behavior in terms of microscopic physics, even in the simplest situation of a nonequilibrium steady state 
(NESS) \cite{noneq,noneq1,fourier0,Bertini,Derrida,Pablo,IFR}. In contrast with equilibrium, the microscopic probability measure associated to a NESS (or mNESS
hereafter) is a complex object, often defined on a fractal support or strange attractor, and is utterly sensitive to microscopic details as the modeling of 
boundary reservoirs (e.g., deterministic vs stochastic) \cite{noneq1,measure0,measure1,GC}. Moreover, the connection between mNESSs and a nonequilibrium analog
of thermodynamic potentials is still murky at best. On the other hand, we do know that essentially different mNESSs (resulting e.g. from different modelings
of boundary baths) describe equally well what seems to be the same macroscopic NESS (or MNESS in short), defined in terms of a few macroscopically smooth fields
\cite{Chernov}. Key for this sort of nonequilibrium ensemble equivalence (which can be formally stated via the \emph{chaotic hypothesis} of Gallavotti and Cohen
\cite{GC}) is the notion of local thermodynamic equilibrium (LTE) \cite{noneq1,Spohn}, i.e. the fact that an interacting nonequilibrium system reaches \emph{locally} 
an equilibrium-like state defined by e.g. a local temperature, density and velocity (the first two related locally via standard thermodynamics), which are 
roughly constant across molecular scales but change smoothly at much larger macroscopic scales, where their evolution is governed by hydrodynamic equations.
LTE plays an important role in physics, being at the heart of many successful theories, from the classical hydrodynamics \cite{Landau} or nonequilibrium 
thermodynamics \cite{deGroot} to recent macroscopic fluctuation theory (MFT) \cite{Bertini,Derrida,Pablo,IFR}. The latter studies dynamic fluctuations of
macroscopic observables arbitrary far from equilibrium, and offers explicit predictions for the large deviation functions (LDFs) controlling the statistics 
of these fluctuations \cite{Touchette}. These LDFs are believed to play in nonequilibrium a role akin to the free energy (or entropy) in equilibrium systems,
establishing MFT as an alternative pathway to derive thermodynamic-like potentials out of equilibrium, bypassing the complexities associated to mNESSs and their
sensitivity to microscopic details \cite{Bertini,Derrida}. The LDFs so obtained exhibit the hallmarks of nonequilibrium behavior, e.g. they are typically nonlocal
(or equivalently nonadditive) \cite{nonlocal} in stark contrast with equilibrium phenomenology. Such nonlocality emerges from tiny, ${\cal O}(N^{-1})$ corrections
to LTE which spread over macroscopic regions of size ${\cal O}(N)$, with $N$ the number of particles in the system of interest \cite{nonlocal}. This shows that
LTE is a subtle property: while corrections to LTE vanish locally in the $N\to \infty$ limit, they have a fundamental impact on nonequilibrium LDFs in the form
of nonlocality, which in turn gives rise to the ubiquitous long-range correlations which characterize nonequilibrium fluids \cite{Bertini,Sengers,longrange}. 

These fundamental insights about LTE and its role out of equilibrium are coming forth from the study of a few oversimplified stochastic models of transport
\cite{Bertini,Derrida,Pablo,IFR,nonlocal}. The question remains however as to whether the emerging picture endures in more realistic systems: Does LTE hold
at the macroscopic level in fluids far from equilibrium? Can we measure corrections to LTE at the microscopic or fluctuating level? Are these corrections the
fingerprints of nonlocality? Here we answer these questions for a quintessential model of a nonequilibrium fluid, the two-dimensional hard-disks system under 
a temperature gradient \cite{Mulero}. 
As we argue below, this model contains the essential ingredients that characterize a large class of fluids whose physics is dominated by the 
short range repulsion between neighboring particles, so we expect some of our results to generalize also to these more realistic model fluids.
In particular, we show below that macroscopic LTE (MLTE) is a very strong property even in the presence of important finite size 
effects, revealing a striking decoupling between the bulk fluid, which behaves  macroscopically, and two boundary layers which sum up all sorts of artificial 
finite-size and boundary corrections to renormalize the effective boundary conditions on the remaining bulk. This bulk-boundary decoupling phenomenon, together
with the robustness of the MLTE property, allows to measure with stunning accuracy the \emph{equilibrium} equation of state (EoS) of hard disks in nonequilibrium
simulations, even across the controversial hexatic and solid phases \cite{hexa}. To search for corrections to LTE, we study the moments of the velocity field and
the total energy. While the former do not exhibit corrections to LTE, the fluctuations of the total energy do pick up these small corrections, the essential
difference coming from the nonlocal character of the higher-order moments of the total energy. This suggests that corrections to LTE are indeed linked to the
nonlocality of the nonequilibrium fluid.

\section{Model and simulation details}
\label{sec2}

One of the simplest ways to simulate a fluid from microscopic dynamics consists in modelling its constituent particles as impenetrable bodies undergoing ballistic motion 
in between ellastic collisions with neighboring particles.
Such oversimplied description picks up however the essential ingredient underlying the physics of a large class of fluids, namely the strong, short-distance 
repulsion between neighboring molecules. This short-range repulsion dominates the local and global emerging structures in the fluid, as well as the nature of interparticle 
correlations \cite{Mulero,Chaikin}. In this way, hard-sphere models and their relatives capture the physics of a large class of complex phenomena, ranging from phase transitions or heat flow to glassy dynamics, jamming, or the physics of liquid crystals and granular materials, to mention just a few 
\cite{Mulero,hexa,Chaikin,soft,glass,deGennes,granular,tails,Alder,Rosenbluth,fourier1,fourier2,shearCL,Pedro,chaos,Risso,Mareschal}, hence defining one of the most successful, 
inspiring and prolific models of physics. 

From a theoretical standpoint, hard-sphere models are also very interesting because 
their low- and moderate-density limits are directly comparable with predictions from kinetic theory (either Boltzmann-type equations  for low-density or ring-kinetic theory for higher densities) \cite{Gass,Resibois}. Moreover, hard-sphere models are commonly used as a reference systems for perturbation approaches to the statistical mechanics of interacting particle systems \cite{Andersen,Hen2,Mulero}. In fact, the idea of representing a liquid by a system of hard bodies can be already found in the work of Van der Waals: his famous equation of state was derived using essentially this principle. In addition, there exist an extensive literature on efficient event-driven molecular dynamics algorithms for hard-sphere models \cite{Marin,Isobe}, which open the door to massive, 
long-time simulations with large numbers of particles. This last feature turns out to be crucial in our case, as 
deviations from LTE are expected to be small and to occur at the fluctuating level \cite{Bertini,Derrida,Pablo,IFR,nonlocal}, 
thus requiring excellent statistics to pick up such a weak signal.

\begin{figure}[t]
\vspace{-0.3cm}
\includegraphics[width=8.5cm]{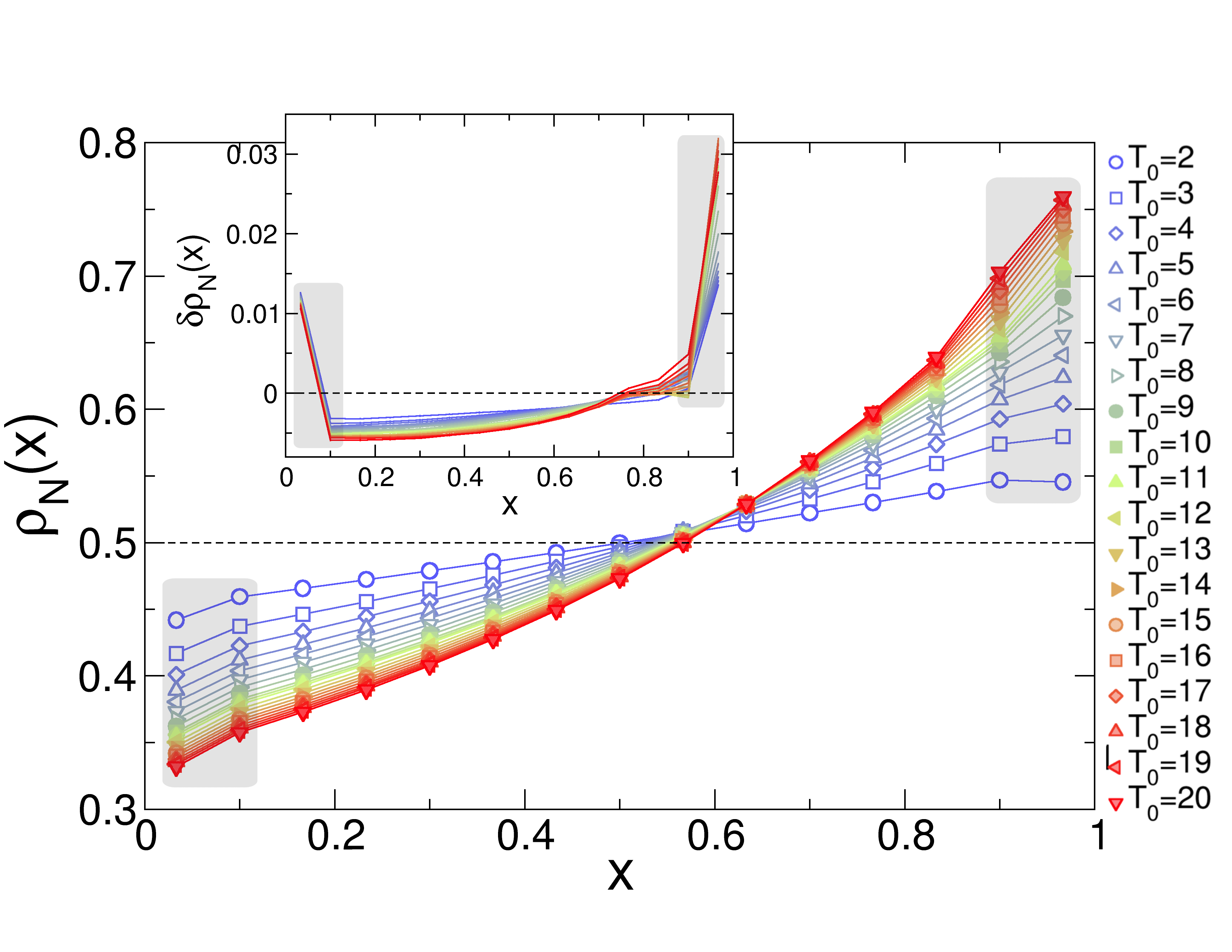}
\vspace{-0.5cm}
\caption{\small (Color online) Density profiles for $N=8838$, $\eta=0.5$ and varying $T_0\in[2,20]$. Shaded areas correspond to boundary layers. Inset: Finite
size effects as captured by $\delta \rho_N(x)\equiv \rho_{N_{\text{max}}}(x) - \rho_{N_{\text{min}}}(x)$, with $N_{\text{max}}=8838$ and $N_{\text{min}}=1456$, for different gradients.  
}
\label{fig1a}
\end{figure}

\begin{figure}[t]
\vspace{-0.3cm}
\includegraphics[width=8.5cm]{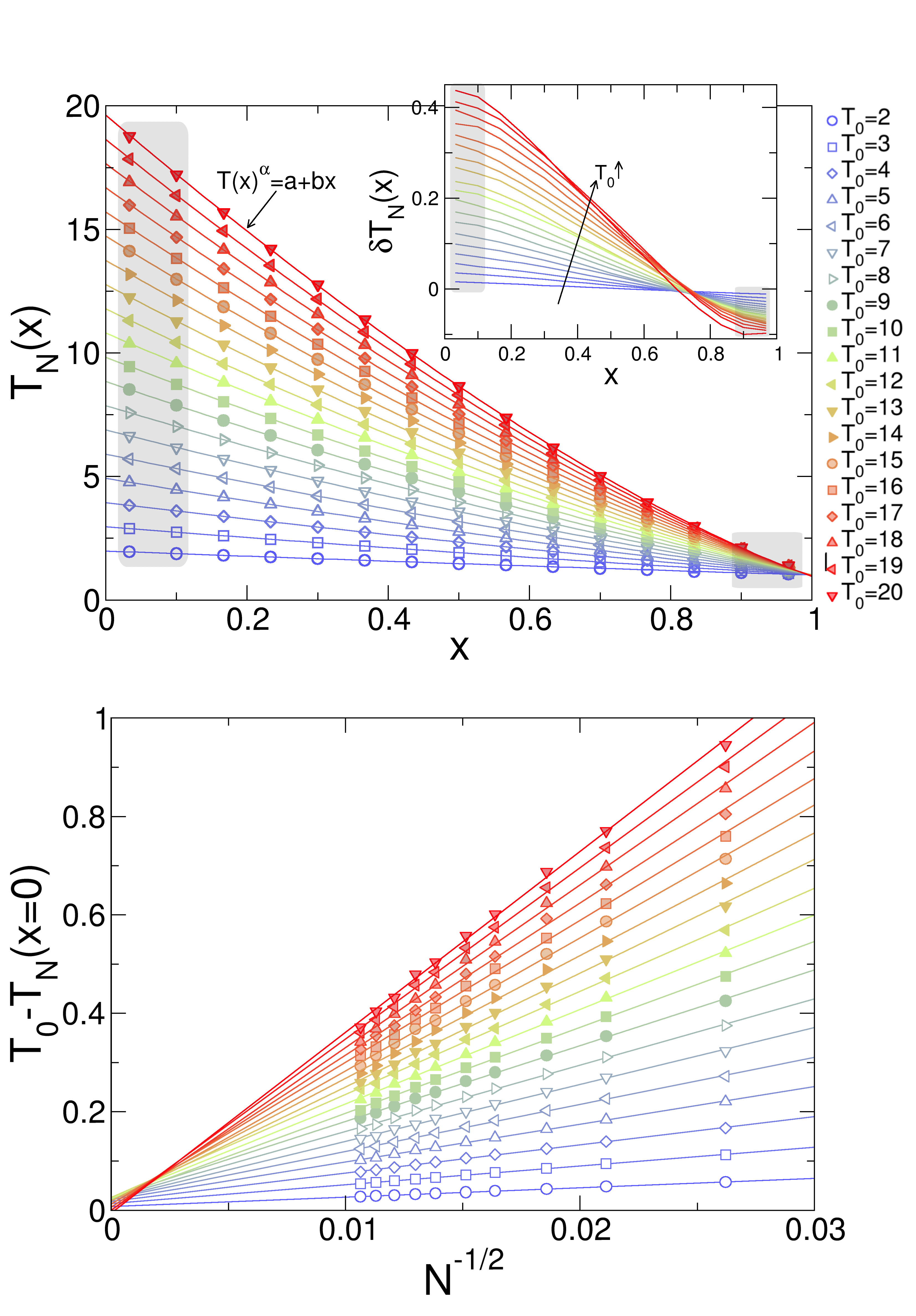}
\vspace{-0.5cm}
\caption{\small (Color online) Top: Temperature profiles for the same conditions as in Fig. \ref{fig1a}, and finite-size effects (inset). Bottom: Temperature
gap at the hot wall vs $1/\sqrt{N}$ for varying $T_0$, and linear fits to the data. 
}
\label{fig1a2}
\end{figure}

In this work we study a system of $N\in [1456,8838]$ hard disks of radius $\ell$ in a two-dimensional box of unit size $L=1$, with stochastic thermal walls
\cite{fourier1,walls,Tenen} at $x=0,~L$ at temperatures $T_0\in[2,20]$ and $T_L=1$, respectively, and periodic boundary conditions along the $y$-direction. The stochastic
thermal walls work as follows: each time a particle hits one of these walls, its velocity $\vv=(v_x,v_y)$ is randomly drawn from a Maxwellian distribution defined by the wall temperature $T_{0,L}$
\ben
\text{Prob}_{0,L}(v_x)&=&\frac{|v_x|}{\sqrt{2\pi T_{0,L}}} \exp{\left(-\frac{v_x^2}{2T_{0,L}}\right)}  \, , \nonumber \\  
\text{Prob}_{0,L}(v_y)&=&\frac{1}{\sqrt{2\pi T_{0,L}}} \exp{\left(-\frac{v_y^2}{2T_{0,L}}\right)} \, , \nonumber
\een
with the additional constraint that the $x$-component of the velocity changes sign. Note that we take units where Boltzmann constant $k_B=1$. The above process simulates in a highly 
efficient manner the flow of energy in and out of the system from an infinite, equilibrium reservoir at the chosen temperature. 
On the other hand, the disks radius is defined as 
\be
\ell=\sqrt{\eta/N\pi} \, ,
\label{diskradius}
\ee
with $\eta=\pi\ell^2N/L^2\in[0.05,0.725]$ the global packing fraction. With these definitions, we can approach the $N\to \infty$ limit at fixed $\eta$ and constant temperature gradient
$\Delta T\equiv|T_L-T_0|/L$. 

In order to characterize the inhomogeneous nonequilibrium steady state in the fluid, we divided the system into 15 virtual cells of linear size 
$\lambda=L/15$ along the gradient direction, and measured locally a number of relevant observables
including the local average kinetic energy, virial pressure, packing fraction, etc., as well as the energy current flowing through the thermal baths and the 
pressure exerted on the walls. Local temperature is then defined via equipartition theorem from the average kinetic energy per particle in each local cell. 
Moreover, in order to look for deviations from LTE, we also measured moments of the velocity field and the total energy, see Section \ref{sec4}.

In order to measure the reduced pressure $Q_N\equiv \pi \ell^2 P_N$ (with $P_N$ the pressure), we used two different methods which yield equivalent results (see below). On  one hand, 
we measured the reduced pressure exerted by the fluid on the thermal walls, $Q_w(N)$, in terms of the average momentum exchanged between the 
colliding particles and the thermal wall per unit length and unit time. On the other hand, we also measured locally in each cell the reduced virial pressure 
$Q_v(x;N)$ defined as
\be
Q_v(x;N)= T_N(x) \rho_N(x)+\frac{\pi \ell^2}{2 \lambda L \tau_{\text{col}}}\la \displaystyle\sum_{\text{col(x)}} \vec{v}_{ij} \cdot \vec{r}_{ij}\ra \, ,
\label{pvirial}
\ee
with $\rho_N(x)$ and $T_N(x)$ the local packing fraction and local temperature, respectively, and where the sum is taken over all collisions 
occurring during a time interval $\tau_{\text{col}}$ in a cell centered at $x$. Here $\vec{v}_{ij}=\vec{v}_i-\vec{v}_j$  is the relative velocity of the 
colliding pair $(i,j)$, 
$\vec{r}_{ij}$ is the vector connecting the particles centers at collision, with $r_{ij}=2\ell$, and angular brackets represent an ensemble average.
In equilibrium, Eq. (\ref{pvirial}) is just the viral pressure \cite{virial}, and our results below show that locally in a nonequilibrium steady state the 
above expression is a sound definition of pressure. In fact, we will show below that both wall and virial definitions of pressure, $Q_w(N)$ and $Q_v(x;N)$, 
yield values consistent with each other and with theoretical predictions. 

Finally, note that time averages of the different observables were performed with measurements every $10$ time units for a total time of $10^6-10^7$ (our time unit was set to one 
collision per particle on average), after a relaxation time of $10^3$ which was sufficient to reach the steady state. Small corrections ($\sim 0.1\%$) due to the spatial 
discretization of density and temperature profiles are explicitly taken into account and subtracted (see Appendix A), and statistical errors in data averages (at a $99.7\%$ 
confidence level) are always plotted.

\section{Macroscopic local equilibrium}
\label{sec3}

Figs.~\ref{fig1a} and \ref{fig1a2} show the density and temperature profiles, $\rho_N(x)$ and $T_N(x)$ respectively, measured for $N=8838$, $\eta=0.5$ and 
varying gradients $\Delta T$. These profiles are in general non-linear, and similar
profiles are measured for different $N$, $\eta$ and $\Delta T$. Interestingly, temperature
profiles in Fig. \ref{fig1a2} (top) can be fitted with high accuracy by the phenomenological law
\be
T(x)^\alpha=ax+b
\label{profT}
\ee
with $\alpha$ an exponent characterizing the apparent nonlinearity. This simple law has been deduced for some two-dimensional Hamiltonian and stochastic models
of heat transport \cite{Eckmann}. In our case, however, the fitted exponent $\alpha$ exhibits a pronounced dependence on $N$ and $\Delta T$ (not shown), with
$\alpha\in[0.681,0.715]$ and no coherent asymptotic behavior, a trait of the strong finite size corrections affecting the hydrodynamic profiles. These corrections
are captured for instance by the finite-size excess density and temperature profiles, defined as
\be
\delta f_N(x)\equiv f_{N_{\text{max}}}(x) - f_{N_{\text{min}}}(x) \, , 
\label{excessdens}
\ee
with $f\equiv \rho,\, T$, and $N_{\text{max}}=8838$ and $N_{\text{min}}=1456$ the maximum and minimum number of particles used in our simulations. The insets in
Figs.~\ref{fig1a} and \ref{fig1a2} show $\delta \rho_N(x)$ and $\delta T_N(x)$ measured for different temperature gradients, signaling the importance of finite-size
corrections in this setting, particularly near the boundaries and most evident for density profiles. Indeed, the thermal walls act as defect lines disrupting the
structure of the surrounding fluid, a perturbation that spreads for a finite penetration length toward the bulk fluid, defining two boundary layers where finite
size effects and boundary corrections concentrate and become maximal, see insets in Figs. \ref{fig1a} and \ref{fig1a2}. Furthermore, the boundary disturbance gives
rise to a thermal resistance or temperature gap between the profile extrapolated to the walls, $T_N(x=0,L)$, and the associated bath temperature $T_{0,L}$. In the
bottom panel of Fig. \ref{fig1a2} we study the system size dependence of this thermal gap, $\gamma_{0,L}(\Delta T,N)\equiv |T_{0,L}-T_N(x=0,L)|$, finding that
\be
\gamma_{0,L}(\Delta T,N) \sim N^{-1/2} \quad \forall \Delta T \, .
\label{gap}
\ee
In this way, the boundary thermal gaps disappear in the thermodynamic limit when approached at constant packing fraction $\eta$ and temperature gradient $\Delta T$.
In order to minimize boundary corrections, we hence proceed to eliminate from our analysis below the boundary layers by removing from the profiles the two cells 
immediately adjacent to each wall, i.e. cells $i=1,2$ and $14,15$ (see shaded areas in Fig. \ref{fig1a}) \cite{Tenen}.

\begin{figure}[t]
\vspace{-0.3cm}
\includegraphics[width=8.5cm]{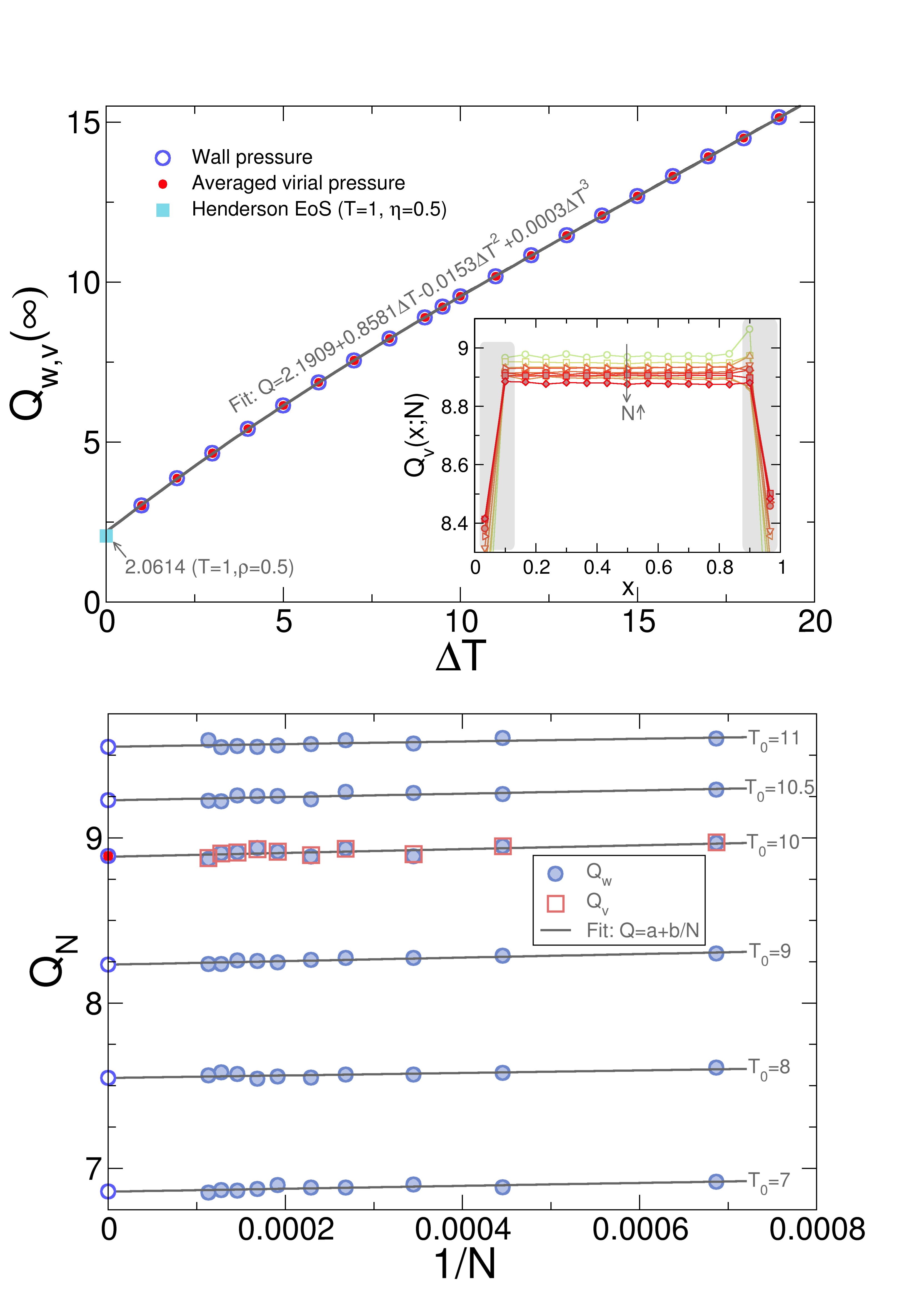}
\vspace{-0.2cm}
\caption{\small (Color online) Top: Bulk-averaged virial and wall reduced pressures as a function of $\Delta T$ for $\eta=0.5$ in the $N\to\infty$ limit, and cubic
fit. The light blue squared symbol ($\blacksquare$) at $\Delta T=0$ represents the equilibrium pressure predicted by the Henderson EoS, see eq. (\ref{Henderson}). 
In both cases (virial and wall pressures) finite-size data scale linearly with $N^{-1}$, see bottom panel. The inset shows virial pressure profiles for different 
$N$, $\eta=0.5$ and $\Delta T=10$, which are constant in the bulk.
}
\label{fig1b}
\end{figure}

As described in Section \ref{sec2}, we also measured the pressure in the nonequilibrium fluid using two different methods, namely by monitoring the (reduced) pressure 
exerted by the fluid on the thermal walls, $Q_w(N)$, and alternatively by measuring a local version of the virial pressure for hard disks, $Q_v(x;N)$, see Eq. (\ref{pvirial}).
The latter yields pressure profiles which are constant across the bulk of the fluid but exhibit a clear estructure at the boundary layers,
see inset in top panel of Fig. \ref{fig1b}. This is of course expected because of the local anisotropy induced by the nearby walls.
A sound definition of the fluid's pressure is then obtained by averaging in the bulk the virial pressure profiles. 
This pressure exhibits strong finite size corrections which scale linearly as $N^{-1}$ for each $\Delta T$, see inset and bottom panel in Fig.
\ref{fig1b}, converging to a well-defined value in the $N\to \infty$ limit. Wall pressures similarly scale as $N^{-1}$, see bottom panel in Fig. \ref{fig1b}, and 
in all cases (both for finite $N$ and in the $N\to\infty$ limit) the measured values agree to a high degree of accuracy and $\forall \Delta T$ with those of the virial 
expression, see top panel in Fig. \ref{fig1b}. 
Notice that the $N\to\infty$ data are consistent with a cubic polynomial dependence of pressure on the temperature gradient, see top panel in Fig. \ref{fig1b}, 
and such cubic dependence is in turn compatible with the equilibrium pressure (for $\Delta T=0$) predicted by the Henderson equation of state \cite{Mulero,Hen}, 
see eq. (\ref{Henderson}) below. 
We hence conclude that both definitions of pressure are compatible with each other and with theoretical predictions, suggesting already a sort of local mechanical equilibrium in the nonequilibrium fluid.

\begin{figure}[t]
\vspace{-0.3cm}
\includegraphics[width=8.5cm]{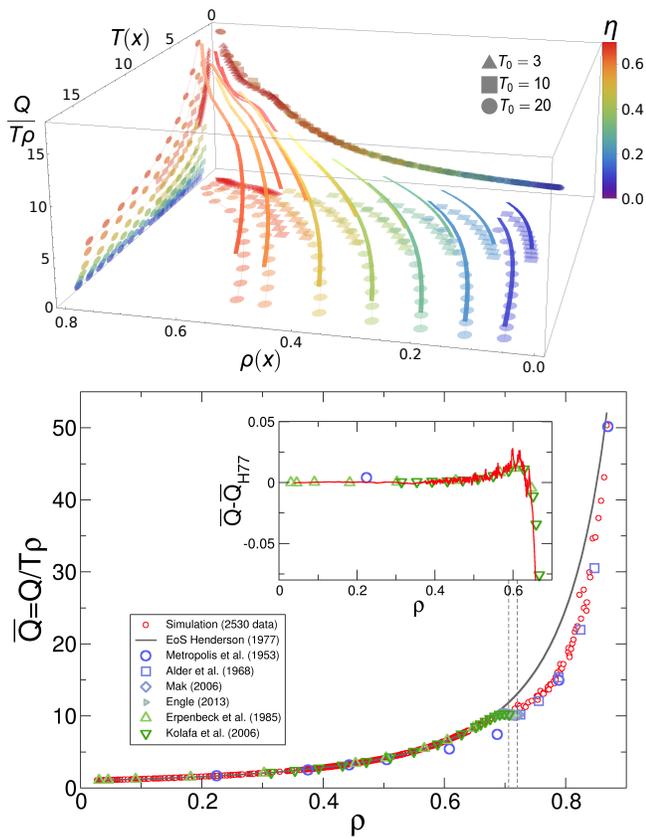}
\vspace{-0.2cm}
\caption{\small (Color online) Top: Compressibility factor $\bar{Q}_N\equiv Q_N/[T_N(x)\rho_N(x)]$ as a function of $T_N(x)$ and $\rho_N(x)$ measured for $N=2900$
and varying $\eta$ and $T_0$. Bottom: $\bar{Q}_N$ vs $\rho_N(x)$ measured for $\eta=0.5$, $N\in [1456,8838]$ and $T_0\in[2,20]$, as well as for $\eta\in[0.05,0.65]$,
$N=2900$ and $T_0=10, 20$, summing up a total of 2530 data points. For comparison, data from previous equilibrium simulations in literature are shown, together with
the Henderson EoS approximation (line). The inset shows a detailed comparison of a running average of our data with equilibrium results after subtracting the leading Henderson behavior. 
}
\label{fig2}
\end{figure}

We now focus on the macroscopic notion of local thermodynamic equilibrium described in the introduction, an issue already explored in early computer simulations \cite{Tenen}. Macroscopic LTE implies that, locally, the density and temperature fields should 
be related via the equilibrium equation of state (EoS), 
\be
Q=\rho T \,\bar{Q}(\rho,T) \, ,
\label{eos}
\ee
where $\bar{Q}(\rho,T)$ is the compressibility factor \cite{Mulero}. Top panel in Fig. \ref{fig2} shows results for $\bar{Q}_N\equiv Q_N/[T_N(x)\rho_N(x)]$ measured
out of equilibrium, as a function of $\rho_N(x)$ and $T_N(x)$. Note that each nonequilibrium simulation, for fixed $(\Delta T,\eta,N)$, covers a fraction of the EoS
surface, thus improving the sampling when compared to equilibrium simulations, which yield a single point on this surface. As hard disks exhibit density-temperature separability 
(i.e. temperature scales out of all thermodynamic relations) \cite{Mulero,Jesus1}, the associated $\bar{Q}$ depends exclusively on density, meaning that a complete collapse is expected for the
projection of the EoS surface on the $\bar{Q}-\rho$ plane, as we indeed observe, see top panel in Fig. \ref{fig2}. Strikingly, although density and temperature profiles,
as well as pressures, all depend strongly on $N$, see Figs. \ref{fig1a}-\ref{fig1b}, the measured $\bar{Q}_N$ as a function of the local density exhibits no finite
size corrections at all, see bottom panel in Fig. \ref{fig2} where a total of 2530 data points for different $N\in[1456,8838]$, $\eta\in[0.05,0.65]$ and $T_0\in[2,20]$
are shown. This strongly suggests a compelling structural decoupling between the bulk fluid, which behaves  macroscopically and thus obeys locally the thermodynamic
EoS, and the boundary layers near the thermal walls, which sum up all sorts of artificial finite-size and boundary corrections to renormalize the effective boundary
conditions on the remaining bulk. This remarkable bulk-boundary decoupling phenomenon, instrumental in the recent discovery of novel scaling laws in nonequilibrium 
fluids \cite{Jesus1,Pablo1d}, is even more surprising at the light of the long range correlations present in nonequilibrium fluids \cite{Bertini,Sengers,longrange}, offering
a tantalizing method to obtain macroscopic properties of nonequilibrium fluids without resorting to unreliable finite-size scaling extrapolations \cite{Pablo1d}. For comparison, 
we include in Fig. \ref{fig2} (bottom) data from several extensive equilibrium simulations carried with different methods during the last 60 years \cite{hexa,simeq},
as well as the Henderson EoS approximation  \cite{Mulero,Hen}
\be
\bar{Q}_{\text{H77}}(\rho)=\left[\frac{1+\rho^2/8}{(1-\rho)^2}-0.043\frac{\rho^4}{(1-\rho)^3} \right] \, ,
\label{Henderson}
\ee
which is reasonably accurate in the fluid phase. The inset in Fig. \ref{fig2} shows a fine comparison of a running average of our data and the equilibrium simulations
in literature, once the leading Henderson behavior has been subtracted. An excellent agreement is found to within $1\%$ relative error, confirming the validity and 
robustness of macroscopic LTE and the bulk-boundary decoupling phenomenon here reported. The accuracy of our data for the EoS is surprising taking into account that 
local cells have at most 500 particles, and many fewer in the typical case. 

\begin{figure}[t]
\vspace{-0.3cm}
\includegraphics[width=7.5cm]{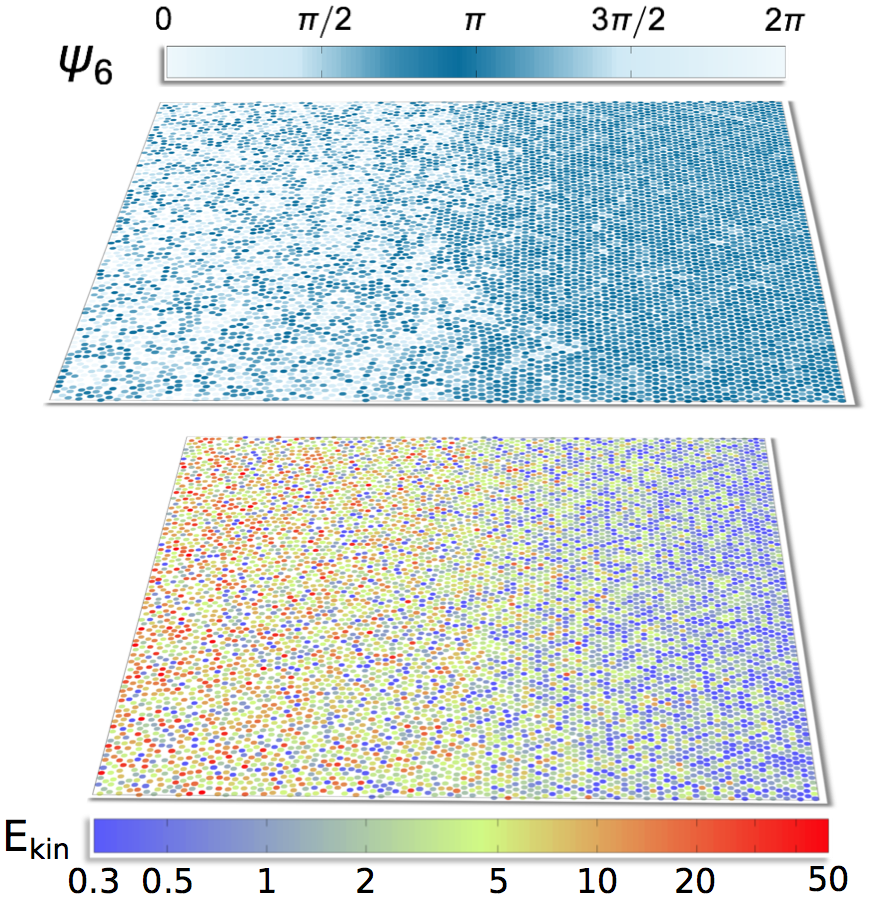}
\vspace{-0.2cm}
\caption{\small (Color online) Snapshot of a typical hard-disks configuration with $N=7838$, $\eta=0.7$ and $T_0=10$ and two color codes representing respectively the
local hexatic order (top) and the kinetic energy (bottom). Inhomogeneous fluid and solid phases coexist for such high $\eta$ under a temperature gradient. 
}
\label{fig2b}
\end{figure}

\begin{figure*}[t]
\vspace{-0.3cm}
\includegraphics[width=12.5cm]{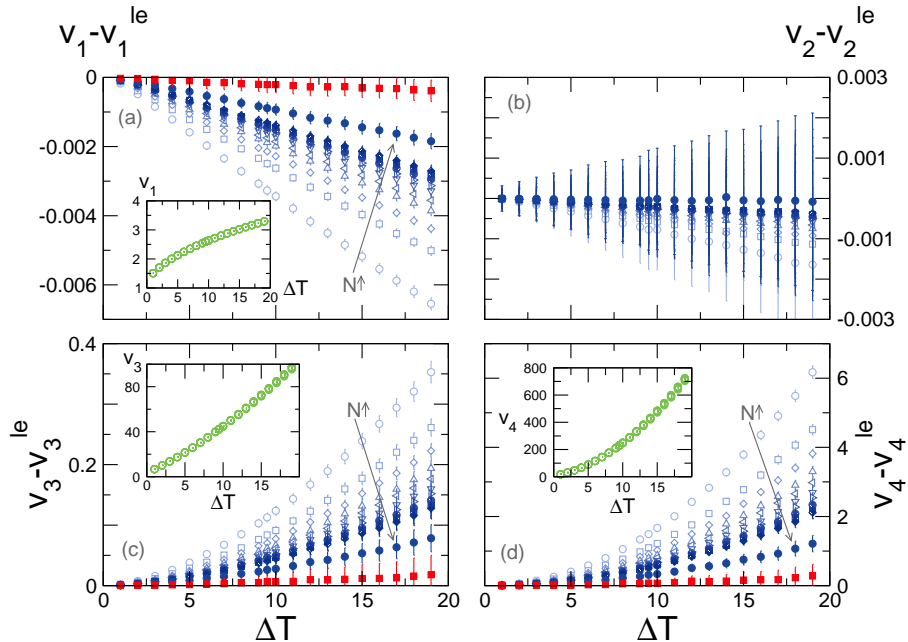}
\vspace{-1cm}
\caption{\small (Color online) Velocity moments after subtracting the LTE contribution, as a function of $\Delta T$ for $\eta=0.5$ and varying $N$. Filled symbols
correspond to the $N\to\infty$ limit before ($\CIRCLE$) and after ($\blacksquare$) correcting the LTE values for the discretization of hydrodynamic profiles and their
fluctuations (see Appendix B). After the correction, no deviations from LTE are observed in velocity moments. The insets show the moments before subtracting the LTE part. 
}
\label{fig3}
\end{figure*}

Note that our data include points across and beyond the controversial liquid-hexatic-solid double phase transition regime \cite{hexa}. In fact, for $\eta\gtrsim 0.6$
a coexistence between a fluid phase near the hot wall and a solid-like phase near the cold one is established. In Fig. \ref{fig2b} we plot a typical configuration in
this nonequilibrium coexistence regime with two color codings. The first one (top) represents the local angular order parameter $\psi_6$ \cite{hexa},  with
\be
\psi_6\equiv \frac{1}{N_b}\sum_{k=1}^{N_b} \text{e}^{\text{i}\phi_k}
\label{fi6}
\ee
where $N_b$ is the number of nearest neighbors of a given particle and $\phi_k$ is the angle of the bond connecting the reference particle with its $k$-neighbor, relative
to an arbitrary direction ($\hat{x}$ in our particular case). The order parameter $\psi_6$ picks the local hexatic order of the symmetry-broken phase, offering an 
interesting method to characterize the interface between the inhomogeneous fluid and solid phases. This coexistence appears in the presence of a strong temperature 
gradient and an associated heat current, as captured by the second color code which represents kinetic energy in the bottom panel of Fig. \ref{fig2b}. This interesting
nonequilibrium fluid/solid coexistence will be investigated in detail in a forthcoming paper \cite{Jesus2}.

\section{Corrections to local equilibrium at the fluctuating level}
\label{sec4}

The macroscopic notion of LTE that we have just confirmed does not carry over however to microscopic scales. The local statistics associated to a fluid's mNESS must
be more complicated than a local Gibbs measure, containing small (but intricate) corrections which are essential for transport to happen. In fact, the local Gibbs
measure in a fluid is even in velocities, while for instance the local energy current is odd, thus leading to a zero average energy current for a mLTE state, in
contradiction with observed behavior. It is therefore the tiny corrections to mLTE that are responsible for transport.

\begin{figure*}[t]
\vspace{-0.3cm}
\includegraphics[width=12.5cm]{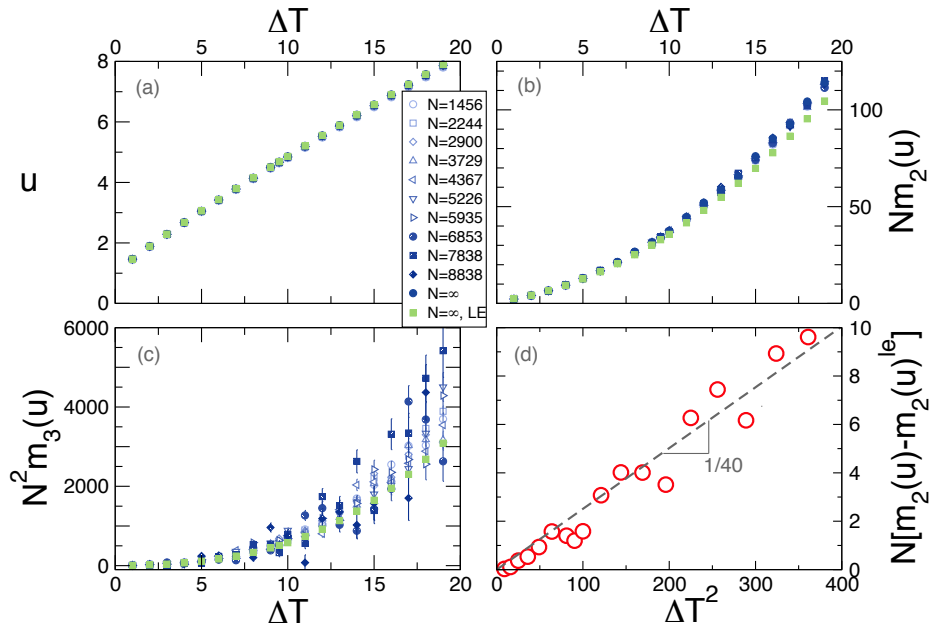}
\vspace{-0.75cm}
\caption{\small (Color online) (a), (b) and (c): Scaled central moments of the total energy as a function of $\Delta T$ for $\eta=0.5$ and varying $N$. Filled symbols correspond 
to the $N\to\infty$ limit of our data ($\CIRCLE$) and after assuming LTE ($\blacksquare$). (d) The excess energy variance (as compared to LTE) scales linearly with $\Delta T^2$
with a slope $\approx 1/40$.  
}
\label{fig4}
\end{figure*}

To search for these corrections, we first measured the local velocity statistics (not shown), which turns out to be indistinguishable within our precision from a
local Maxwellian distribution with the associated local hydrodynamic fields as parameters. Experience with stochastic lattice gases \cite{nonlocal,Spohn} suggests
that deviations from LTE are more easily detected in global observables, so we also measured the global velocity moments 
\be
v_n\equiv \la \frac{1}{N} \sum_{i=1}^N | \vv_i|^n \ra \, ,
\label{vmom}
\ee
with $n=1,2,3,4$, and compared them with the LTE predictions based on a local Gibbsian measure, 
\be
v_n^{\text{le}}\equiv \frac{a_n}{\eta}\int_{0}^1 dx \rho(x)T(x)^{n/2} \, ,
\label{vmomLE}
\ee
see Appendix B for a derivation, with $a_1=\sqrt{\pi/2}$, $a_2=2$, $a_3=3\sqrt{\pi/2}$ and $a_4=8$. The LTE corrections we are looking for are tiny, so several
correcting factors must be taken into account, e.g. the effect of the discretized hydrodynamic profiles and their fluctuations on $v_n^{\text{le}}$, and the appropriate
error propagation, see Appendix B. Fig. \ref{fig3} shows the first four velocity moments as a function of $\Delta T$, after subtracting the (uncorrected) LTE
contribution, as well as the $N \to \infty$ extrapolation before and after the corrections to $v_n^{\text{le}}$ mentioned above. Although a naive comparison of
velocity moments with raw LTE estimates would suggest that LTE already breaks down at this level, it becomes apparent that, after properly accounting for all
corrections mentioned, no deviations from LTE are observed in velocity moments.

To further pursue the analogy with stochastic lattice gases \cite{nonlocal,Spohn}, we also studied the central moments $m_n(u)\equiv \la (u - \la u\ra)^n\ra$ of the fluid's total energy per particle, 
\be
u\equiv \frac{1}{N}\sum_{i=1}^N \frac{1}{2}m \vv_i^2 \, .
\label{ener}
\ee
Figs. \ref{fig4}.a-c show the measured $m_n(u)$, $n=1,2,3$, as a function of $\Delta T$ for $\eta=0.5$ and different $N$, together with the $N\to \infty$ extrapolation
of our data and the (corrected) LTE estimates for energy moments, $m_n(u)^{\text{le}}$ (see Appendix C). We observe that, while the average energy does indeed follow 
the LTE behavior, $\la u\ra\sim \la u\ra^{\text{le}}$, energy fluctuations (as captured by $m_2(u)$ in Fig. \ref{fig4}.b) exhibit increasing deviations from the
LTE estimate. In fact, the excess energy fluctuations scale linearly with the squared gradient,
\be
\delta m_2(u)\equiv N[m_2(u)-m_2(u)^{\text{le}}] \approx +\frac{1}{40} \Delta T^2 \, ,
\label{udt2}
\ee
see Fig. \ref{fig4}.d, a result strongly reminiscent of the behavior observed in schematic models like the Kipnis-Marchioro-Presutti (KMP) model of heat transport
or the symmetric exclusion process (SEP) \cite{nonlocal}, where $\delta m_2(u)=\pm \Delta T^2/12$. Interestingly, energy fluctuations for hard disks are enhanced
with respect to LTE, $\delta m_2(u)>0$, as happens for the KMP model and contrary to the observation for SEP \cite{nonlocal}, although the excess amplitude is roughly
three times smaller for disks. A natural question is then why we detect corrections to LTE in energy fluctuations but not in velocity moments.
The answer 
lies in the nonlocal character of energy fluctuations. In fact, while $v_2$ includes only a sum of local factors, $m_2(u)$ includes nonlocal contributions of the form
$\la \vv_i^2 \vv_j^2\ra$, $i\neq j$, summed over the whole system. Small,  ${\cal O}(N^{-1})$ corrections to LTE extending over large, ${\cal O}(N)$ regions give
rise to weak long-range correlations in the system which, when summed over macroscopic regions, yield a net contribution to energy fluctuations, which thus depart
from the LTE expectation \cite{nonlocal,Bertini,Derrida}. In this way, the observed breakdown of LTE at the energy fluctuation level is a reflection of the nonlocality
of the underlying large deviation function governing fluctuations in the nonequilibrium fluid. This result is, to our knowledge, the first evidence of a nonlocal LDF
in a realistic model fluid, and suggests to study in more detail large deviation statistics in hard disks.

\begin{figure*}[t]
\vspace{-0.3cm}
\hspace{-3.8cm}
\includegraphics[width=16.cm]{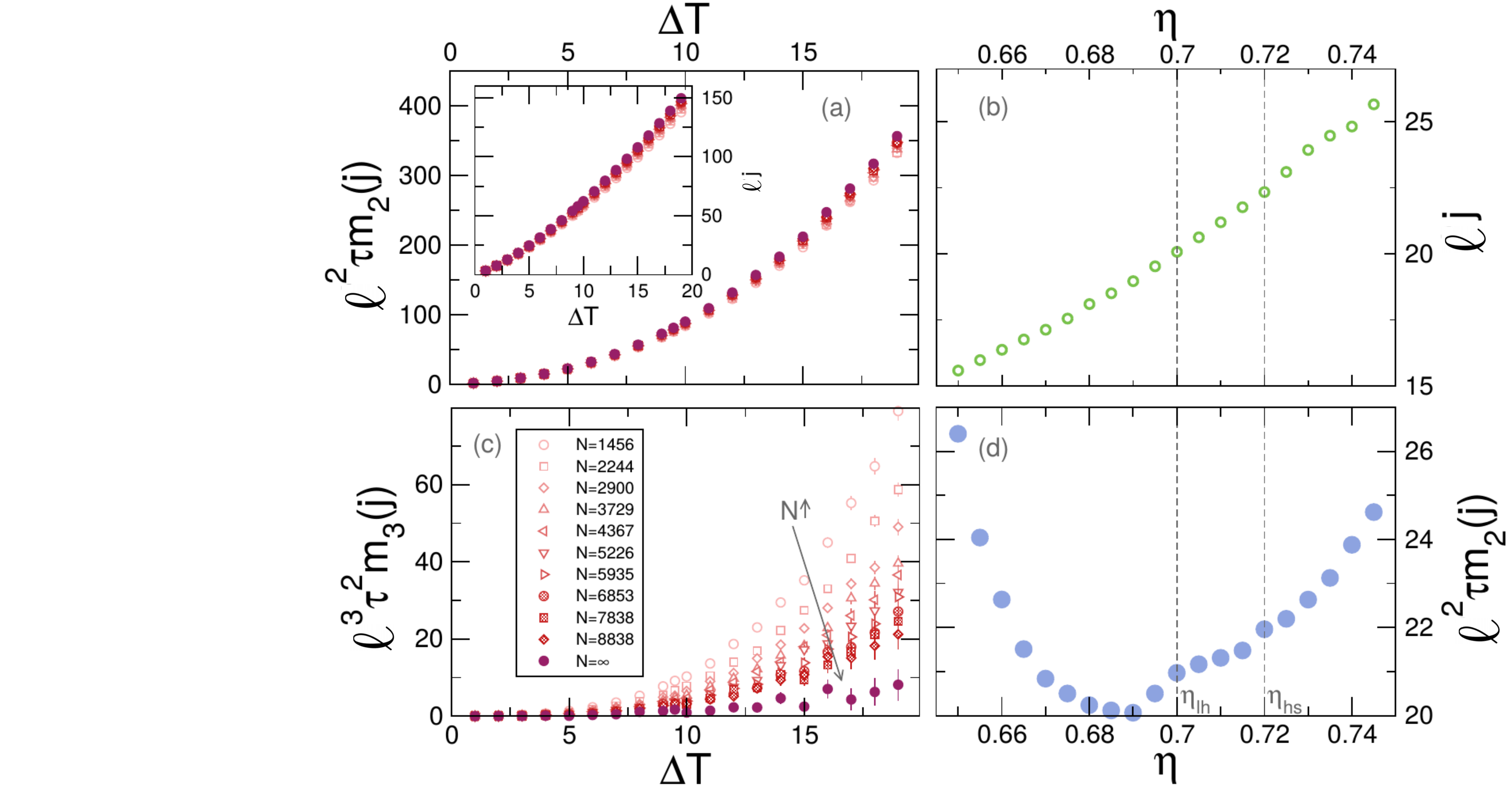}
\vspace{-0.1cm}
\caption{\small (Color online) (a), (c) and inset: Scaled central moments of the time-averaged current vs $\Delta T$ for $\eta=0.5$ and varying $N$. (b), (d) Average current and its variance as
a function of the global packing fraction for $\Delta T=10$ and $N=8838$. The liquid-to-hexatic ($\eta_{\text{lh}}$) and hexatic-to-solid ($\eta_{\text{hs}}$) transition points are signaled. 
}
\label{fig5}
\end{figure*}

A first step in this direction consists in investigating the statistics of the energy current flowing through the system during a long time $\tau$, a key
observable out of equilibrium. In order to do so, and given the difficulties in sampling the tails of a LDF, we measured the central moments of the
time-integrated energy current for $\tau=10$. Figs. \ref{fig5}.a,c show the behavior of the first three central moments of the current as a function of the temperature gradient, while Figs. \ref{fig5}.b,d 
show the dependence of the first and second current moments on the global packing fraction $\eta$. 
Notice that we scale the current by the radius of the particles $\ell$ in order to normalize data for different number of particles. 
A first observation is that our measurements for the current third moment seem compatible with zero $\forall \, \Delta T$ in the
asymptotic $N\to \infty$ limit, See Fig. \ref{fig5}.c, suggesting Gaussian current fluctuations within our accuracy level.
On the other hand it is interesting to note that, while the average current (as well as the total energy moments --not shown--) are smooth functions
of the global packing fraction, see Fig. \ref{fig5}.b, current fluctuations as captured by $m_2(j)$ exhibit a remarkable structure, with a minimum right before the liquid-to-hexatic equilibrium
transition $\eta_{\text{lh}}\approx 0.7$, where a first inflection point appears, followed by another one at the hexatic-to-solid transition $\eta_{\text{hs}}\approx 0.72$, see panel (d) in Fig. \ref{fig5}. This shows that
hints of the double phase transition arise in the transport properties of the nonequilibrium hard-disk fluid, a behavior that deserves further exploration \cite{Jesus2}.

\section{Discussion}
\label{sec5}

In summary, we have probed LTE in a quintessential model fluid, the hard-disks system, finding that macroscopic LTE is a very robust property.
This is so even under strong finite-size effects, due to a remarkable bulk-boundary structural decoupling by which all sorts of finite-size and boundary corrections are renormalized into new boundary conditions for the bulk fluid, which in turn obeys the macroscopic laws. We use these properties to measure with high accuracy the hard-disks EoS, even across the fluid-hexatic-solid transition regime. However, weak but clear violations of microscopic or statistical LTE are found in the fluctuations of the total energy which strongly suggest that the nonequilibrium potential governing the driven fluid's macroscopic behavior is intrinsically non-local. It would be therefore interesting to investigate more in depth the large deviation statistics of hard disks, using both macroscopic fluctuation theory and simulations of rare events.

\acknowledgments
Financial support from Spanish projects FIS2009-08451(MICINN) and FIS2013-43201-P (MINECO), NSF Grant DMR1104500, University of Granada, Junta de Andaluc\'{\i}a projects P06-FQM1505, P09-FQM4682 and GENIL PYR-2014-13 project is acknowledged.

\appendix

\onecolumngrid

\section{Corrections due to discretization effects in density and temperature profiles}
\label{apA}

We measure in the steady state the local temperature (i.e. local average kinetic energy) and local packing fraction at each of the $15$ cells in which we divide the simulation box along the gradient (i.e. $x$-) direction. In order to minimize cells boundary effects, when a disk overlaps with any of the imaginary lines separating two neighboring cells, it contributes to the density and kinetic energy of each cell proportionally to its overlapping area. Note that the number of cells is constant in all simulations, independently of $N$, $\eta$, $T_0$ or $T_L$, so each cell becomes \emph{macroscopic} asymptotically as $N\to\infty$. We now relate averages around a finite neighborhood of a given point in space with the underlying continuous profiles in order to subtract any possible bias or systematic correction from the data. In particular, let's $T_C$ and $\rho_C$ be the temperature and packing fraction in a cell centered at $x_c\in[0,L]$ of size $\Delta$. Assuming that there exist continuous (hydrodynamic) profiles $T(x)$ and $\rho(x)$, we can relate the cell averages to the continuous profiles by noting that
\ben
T_C&=&\frac{1}{\Delta\rho_C}\int_{x_c-\Delta/2}^{x_c+\Delta/2}dx\,\rho(x)T(x) \, , \nonumber \\
\rho_C&=&\frac{1}{\Delta}\int_{x_c-\Delta/2}^{x_c+\Delta/2}dx\,\rho(x) \, . \nonumber 
\een
We may expand now the continuous profiles around $x_c$ inside the cell of interest and solve the above integrals. Keeping results up to $\Delta^2$ order we arrive at
\begin{eqnarray}
T_C&=&\frac{1}{\rho_C}\biggl[\rho(x_c)T(x_c)+\frac{\Delta^2}{24}\frac{d^2}{dx^2}\left[\rho(x)T(x)\right]_{x=x_c}+O(\Delta^3)\biggr]\, , \nonumber\\
\rho_C&=&\rho(x_c)+\frac{\Delta^2}{24}\frac{d^2\rho(x)}{dx^2}\vert_{x=x_c}+O(\Delta^3) \, .\nonumber
\end{eqnarray} 
By inverting the above expressions, we obtain the desired result, namely
\begin{eqnarray}
T(x_c)&=&T_C-\frac{1}{24}\biggl[\frac{2}{\rho_C}(\rho_{C+1}-\rho_C)(T_{C+1}-T_C) + T_{C+1}-2T_C+T_{C-1} \biggr]\label{TC} \, ,\\
\rho(x_c)&=&\rho_C-\frac{1}{24}\left[\rho_{C+1}-2\rho_C+\rho_{C-1}\right] \, ,\label{rhoC}
\end{eqnarray}
which yields the points of the underlying continuous profiles, $T(x_c)$ and $\rho(x_c)$, in terms of the measured observables, $T_C$ and $\rho_C$ respectively. These corrections to the cell density and temperature are typically small ($\sim 0.1\%$), though important to disentangle the different finite-size effects in order to obtain the striking collapse of the hard-disks EoS described in the main text.

\section{Velocity moments under local equilibrium and corrections}
\label{apB}

The local equilibrium probability measure to observe a particle configuration with positions $\vec{r}_i$ and momenta $\vec{p}_i$, $\forall i\in [1,N]$, can be written as \cite{Spohn}
\begin{equation}
\mu^{\text{le}}(\vec{r}_1,\ldots,\vec{r}_N;\vec{p}_1,\ldots,\vec{p}_N)\propto \exp\left\{-\sum_{i=1}^N\beta(\epsilon x_i)\left[ \frac{{\vec{p}_i}\,^2}{2m}+\frac{1}{2}\sum_{j\neq i}\Phi(\vec{r}_i-\vec{r}_j) \right]\right\} \, ,
\label{muLE}
\end{equation}
where $\epsilon$ is the parameter that connects the microscopic scale  with the hydrodynamic one, $\beta(\epsilon x_i)$ is the local inverse temperature around $x_i$, and $\Phi(\vec{r})$ is the interparticle potential. Note that we have already assumed that temperature varies only along the $x$-direction. Under this microscopic LTE hypothesis, one may argue that the probability density for any particle to have a velocity modulus equal to $v$ is given by
\begin{equation}
f(v)=\frac{v}{\eta} \int_{0}^1 dx \frac{\rho(x)}{T(x)}\exp\left[-\frac{v^2}{2T(x)} \right] \, ,
\end{equation} 
with $\eta=\int_{0}^1dx \rho(x)$ the global packing fraction, $\rho(x)$ being its local version. Then, by definition, the local equilibrium velocity moments follow as
\begin{equation}
v_n^{\text{le}}\equiv\langle v^n\rangle_{\text{le}}=\frac{a_n}{\eta}\int_{0}^1 dx \rho(x)T(x)^{n/2}\label{vnle}
\end{equation}
where $a_1=(\pi/2)^{1/2}$, $a_2=2$, $a_3=3(\pi/2)^{1/2}$ and $a_4=8$. Now, in order to compare the measured velocity moments in the main text with the local equilibrium expectations, $v_n^{\text{le}}$, we first have to express the latter in terms of our observables. For that, three steps should be taken, namely:
\begin{itemize}
\item Write eq. (\ref{vnle}) as a function of the measured cell temperatures and densities, $T_C$ and $\rho_C$, following eqs. (\ref{TC})-(\ref{rhoC}).
\item Take into account the fact that both $T_C$ and $\rho_C$ have errorbars and thus should be considered as fluctuating variables around their average. Note also that the fluctuations of $\rho(x)$ should be constrained to a constant average $\eta$.
\item Consider the correct error propagation to compute the errorbars  of $v_n^{\text{le}}$.
\end{itemize}

\subsection{Conversion to cell variables}
We now express eq. (\ref{vnle}) in terms of cell variables. In order to proceed, we write
\begin{equation}
\tilde v_n^{\text{le}}=\frac{a_n}{\eta}\sum_C\int_{x_c-\Delta/2}^{x_c+\Delta/2}dx \rho(x)T(x)^{n/2} 
\end{equation}
where $x_c$ are the center of the cells. We may expand the previous expression up to order $\Delta^2$ to arrive at
\begin{eqnarray}
\tilde v_n^{\text{le}}&=&\frac{a_n}{\eta}\Delta\sum_C\Biggl[\rho(x_c)T(x_c)^{n/2}+\frac{\Delta^2}{24}\biggl[\frac{d^2\rho(x)}{dx^2}\biggr\vert_{x=x_c}T(x_c)^{n/2}
+n\frac{d\rho(x)}{dx}\biggr\vert_{x=x_c}
T(x_c)^{n/2-1}\frac{dT(x)}{dx}\biggr\vert_{x=x_c}\nonumber\\
&+&\frac{n}{2}(\frac{n}{2}-1)\rho(x_c)T(x_c)^{n/2-2}
\left(\frac{dT(x)}{dx}\biggr\vert_{x=x_c}\right)^2+\frac{n}{2}\rho(x_c)T(x_c)^{n/2-1}\frac{d^2T(x)}{dx^2}\biggr\vert_{x=x_c}
\biggr]
\Biggr] \, ,
\end{eqnarray}
and using eqs. (\ref{TC})-(\ref{rhoC}), we obtain
\begin{equation}
\tilde v_n^{\text{le}}=\frac{\Delta a_n}{\eta}\sum_C\rho_CT_C^{n/2}+\frac{\Delta a_n}{96\eta}n\left(n-2\right)
\sum_C\rho_CT_C^{n/2-2}\left(T_{C+1}-T_C\right)^2\label{vnle2}
\end{equation}

\subsection{Accounting for cell temperature and density fluctuations} 
Due to the finite number of measurements in simulations, any magnitude and, in particular, the cell observables will exhibit fluctuations that will affect the observed averaged behavior. We now assume that these fluctuations are Gaussian and obey
\begin{eqnarray}
\rho_C&=&\bar\rho_C+\gamma_C\xi_C\, , \quad\quad \text{Prob}(\xi_1,\ldots,\xi_{M})=\sqrt{2\pi M}\prod_{i=1}^{M}\left[\frac{1}{\sqrt{2\pi}}e^{-\xi_i^2/2}\right]\delta\left(\sum_{i=1}^{M}\xi_i\right)\nonumber\\
T_C&=&\bar T_C+\sigma_C\zeta_C\, ,\quad\quad \text{Prob}(\zeta_1,\ldots,\zeta_{M})=\prod_{i=1}^{M}\left[\frac{1}{\sqrt{2\pi}}e^{-\zeta_i^2/2}\right]\label{fluc}
\end{eqnarray}
where $M=15$ is the number of cells, $\gamma_C$ and $\sigma_C$ are the measured (empirical) errors associated to the average density and temperature at cell $C$, $\bar\rho_C$ and $\bar T_C$ respectively, and $\xi_C$ and $\zeta_C$ are Gaussian random variables with zero mean and variance one. Notice that the density noise takes into account the fact that the total density is constant. For the case studied here we may assume that the errors are ${\cal O}(\Delta)$ and we will expand results up to orders $\sigma^2$ or $\gamma^2$. Now, by substituting $\rho_C$ and $T_C$ in eq. (\ref{vnle2}) by the fluctuating expressions in (\ref{fluc}) and then averaging over the noise distributions, we arrive at
\begin{equation}
v_n^{\text{le}}\equiv\langle \tilde v_n^{\text{le}}\rangle_{\xi,\zeta}=\frac{\Delta a_n}{\eta}\sum_C \bar\rho_C\langle T_C^{n/2}\rangle_{\zeta}+\frac{\Delta a_n}{96\eta}n\left(n-2\right)
\sum_C\bar\rho_C\bar T_C^{n/2-2}\left(\bar T_{C+1}-\bar T_C\right)^2+O(\Delta^3)\label{vnle3}
\end{equation}
where
\begin{equation}
\langle T_C^{n/2}\rangle_{\zeta}=\bar T_C^{n/2}+\frac{1}{8}n(n-2)\bar T_C^{n/2-2}\sigma_C^2+O(\sigma_C^4)
\end{equation}
It therefore becomes clear that fluctuations in each cell add a small correction to $v_n^{\text{le}}$.

\subsection{Computing errorbars in the local equilibrium approximation} 
Once we know how to compute $v_n^{\text{le}}$ from our set of data, we want to obtain their errorbars, $\chi_n$, defined as
\begin{equation}
\chi_n^2=\langle  (\tilde v_n^{le})^2\rangle_{\xi,\zeta}-\langle \tilde v_n^{le}\rangle_{\xi,\zeta}^2
\end{equation}
In order to do so, we first need to know that
\begin{equation}
\langle \xi_C^2\rangle_{\xi}=1-\frac{1}{M}\quad,\quad\langle \xi_C\xi_{C'}\rangle_{\xi}=-\frac{1}{M}\quad \text{if}\quad C\neq C' \, .
\end{equation}
It is then easy to show that
\begin{equation}
\chi_n^2=\left(\frac{\Delta a_n}{\eta}\right)^2\left[\sum_{C}\left(b_n\sigma_C^2\bar\rho_C^2\bar T_C^{n-2}+\gamma_C^2\bar T_C^n\right)-\frac{1}{M}\left(\sum_C\gamma_C\bar T_C^{n/2} \right)^2\right]
\end{equation}
with $b_1=1/4$, $b_2=1$, $b_3=9/4$ and $b_4=4$. To better appreciate the relevance of the above corrections, it is useful to define
\begin{equation}
({v_n}^{\text{le}})_0\equiv\frac{\Delta a_n}{\eta}\sum_C \bar\rho_C \bar T_C^{n/2}\quad,\quad ({v_n}^{\text{le}})_{\text{corr}}\equiv {v_n}^{\text{le}}-({v_n}^{\text{le}})_0
\end{equation}
That is, $({v_n}^{\text{le}})_{\text{corr}}$ contains the $\Delta^2$ effects due to the finite sizes of the boxes and the effects of the errors on the measured temperature and density profiles. Though small, this correction term is essential to confirm that corrections to LTE does not show up in our measurements of velocity moments, see main text.

\section{Energy central moments under the local equilibrium hypothesis}
\label{apC}

To compute the local equilibrium expressions for the central moments of the total energy per particle let us start with the equilibrium expression for the energy fluctuations in the grand canonical ensemble
\begin{equation}
m_n(U)=(-1)^{n+1}\frac{\partial^n}{\partial\beta^n}\ln\Xi
\end{equation}
where $U$ is the system total energy, see the Hamiltonian $H_N$ defined in eq. (\ref{muLE}) above, $m_n(U)\equiv\la(U-\la U\ra)^n \ra$, and
\begin{equation}
\Xi=\sum_{N=0}^{\infty} z^NZ_N\quad,\quad Z_N= \frac{1}{N!h^{2N}}\int_S d\vec r_N\int d\vec p_Ne^{-\beta H_N}
\end{equation}
For hard disks the canonical partition function $Z_N$ can be written:
\begin{equation}
Z_N=\frac{(2\pi)^N}{N!h^{2N}\beta^N}\Lambda(N,S,r)
\end{equation}
where $\Lambda$ is the configurational part of the canonical partition function which does not depend in this case on temperature. The energy fluctuations then follow as
\begin{eqnarray}
m_2(U)&=&T^2\left[\langle N\rangle+m_2(N) \right]\nonumber\\
m_3(U)&=&T^3\left[2\langle N\rangle+3 m_2(N)+m_3(N) \right] \,
\end{eqnarray}
with $m_n(N)$ the central moments of the total number of particles in the grand canonical ensemble. Assuming now that local equilibrium holds we can write
\begin{eqnarray}
\langle U\rangle^{\text{le}}&=&\int_0^1 dx \langle N_x\rangle T(x) \, , \nonumber\\
m_n(U)^{\text{le}}&=&\int_0^1 dx \, m_n(U_x) \, ,
\end{eqnarray}
where $N_x$ ($U_x$) is the local number of particles (local energy) at position $x$. In this way, if $N$ is the total number of particles in the system, we arrive at
\begin{eqnarray}
u^{\text{le}}&=& \frac{\langle U\rangle}{N}=\frac{1}{\eta} \int_{0}^1 dx \rho(x) T(x) \, , \nonumber\\
Nm_2(u)^{\text{le}}&=&\frac{m_2(U)^{\text{le}}}{N}=\frac{1}{\eta} \int_{0}^1 dx \rho(x)T(x)^2\left[1+\frac{m_2(N_x)}{\langle N\rangle_x} \right] \, , \nonumber\\
N^2m_3(u)^{\text{le}}&=&\frac{m_3(U)^{\text{le}}}{N}=\frac{1}{\eta} \int_{0}^1 dx \rho(x) T(x)^3\left[1+\frac{3}{2}\frac{m_2(N_x)}{\langle N\rangle_x}+\frac{1}{2}\frac{m_3(N_x)}{\langle N\rangle_x} \right] \, .
\end{eqnarray}
These expressions define the local equilibrium prediction for the first 3 central moments of the total energy per particle in our hard disk system. Note that the calculation of the corrections to these LTE estimates due to the discreteness of the measured profiles is similar to that explained in Appendix \ref{apB} for the velocity moments.

\end{document}